\documentclass[
aps,pra,
reprint,
superscriptaddress,
amsmath,
amssymb
]{revtex4-1}

\usepackage[utf8]{inputenc}
\usepackage{times}
\usepackage{microtype}
\usepackage{graphicx}
\usepackage{upgreek}
\usepackage{bm}
\usepackage{hyperref}
\usepackage{xcolor}
\usepackage[normalem]{ulem}
\usepackage{placeins}
\usepackage{amsmath}
\usepackage{dcolumn}
\usepackage{comment}
\usepackage{amsthm}

\begin{document}

\title{Quantum games and quantum strategies}
\author{Jens Eisert}

\address{
Institut f{\"u}r Physik, Universit{\"a}t Potsdam,
14469 Potsdam,
Germany}

\author{Martin Wilkens}

\address{
Institut f{\"u}r Physik, Universit{\"a}t Potsdam,
14469 Potsdam,
Germany}

\author{Maciej Lewenstein}
\address{
Institut f{\"u}r Theoretische Physik, Universit{\"a}t Hannover,
30167 Hannover, Germany
}
\date{\today}

\begin{abstract}
We investigate the quantization of non-zero sum games. For the particular case of the Prisoners' Dilemma we show that this
game ceases to pose a dilemma if quantum strategies are allowed for. We also construct a particular quantum strategy which always gives reward if played against any classical strategy. 
\end{abstract}
\maketitle

One might wonder what games and physics could have possibly in common.
After all, games like chess or poker seem to heavily rely on bluffing,
guessing and other activities of unphysical character. Yet, as was
shown by von Neumann and Morgenstern \cite{NeuMor47}, conscious choice
is not essential for a theory of games. At the most  
abstract level, game theory is about numbers that entities are  
efficiently acting to maximize or minimize \cite{Mye91}. For a quantum 
physicist  
it is then legitimate to ask what happens if linear superpositions  
of these actions are allowed for, that is if games are generalized  
into the quantum domain.

There are several reasons why quantizing games may be interesting.
First, classical game theory is a well established discipline of
applied mathematics \cite{Mye91} which has found numerous applications
in economy, psychology, ecology and biology
\cite{Mye91,DawXX}. Since it is based on
probability to a large extend, there is a fundamental interest in
generalizing this theory to the domain of quantum probabilities.  
Second, if the ``Selfish Genes'' \cite{DawXX} are
reality, we may speculate that games of survival
are being played already on the molecular level where quantum
mechanics dictates the rules. 
Third, there is an intimate connection between 
the theory of games and the theory of quantum communication.
Indeed, whenever a player passes
his decision to the other player or the game's arbiter, 
he in fact communicates information, 
which -- as we live in a quantum world --
is legitimate to think of as quantum
information. On the other hand it has recently been  
transpired that eavesdropping in quantum-channel communication
\cite{BenBra84,Eke91,GisHut97} and optimal cloning \cite{Wer98} can
readily be conceived a strategic game between two or more players, the
objective being to obtain as much information as possible in a given
set-up. Finally, quantum mechanics may well be  
useful to win some specially designed zero-sum unfair games, like PQ  
penny flip, as was recently demonstrated by Meyer \cite{Mey99},
and it may assure fairness in remote gambling \cite{Vaidman}.

In this letter we consider non-zero sum games where -- in contrast  
to zero-sum games -- the two players no longer appear in strict  
opposition to each other, but may rather benefit from mutual  
cooperation. A particular instance of this class of games, which has  
found widespread applications in many areas of science,  is the  
Prisoners' Dilemma. In the Prisoners' Dilemma, each of the two  
players, Alice and
Bob, must independently decide whether she or he chooses to defect
(strategy $D$) or cooperate (strategy $C$). Depending on their  
decision taken, each player receives a certain pay-off -- see Tab.\
\ref{tab:PayOff}. The objective of each player is to maximize his or
her individual pay-off. The catch of the dilemma is that $D$ is the
{\it dominant strategy}\/, that is, rational reasoning forces each
player to defect, and thereby doing substantially worse than if they
would both decide to cooperate \cite{Footnote1}. In terms of game
theory, mutual defection is also a {\it Nash equilibrium}
\cite{Mye91}: in contemplating on the move $DD$ in retrospect, each of
the players comes to the conclusion that he or she could not have done
better by unilaterally changing his or her own strategy
\cite{Footnote2}.

In this paper we give a physical model of the Prisoners' Dilemma,
and we show that -- in the context of this model -- the players escape
the dilemma if they both resort to quantum strategies.
Moreover, we shall demonstrate that (i) there exists a particular pair
of quantum strategies which always gives reward and is a Nash
equilibrium and (ii) there exist a particular quantum strategy which
always gives at least reward if played against any classical strategy.

The physical model consists of (i) a source of two  
bits, one bit for each player, (ii) a set of physical 
instruments which enable the player to
manipulate his or her own bit in a strategic manner, and (iii) a
physical measurement device which determines the players' pay-off from
the state of the two bits. All three ingredients, the  
source, the players' physical instruments, and the pay-off physical  
measurement device are assumed to be perfectly known to both  
players.

\begin{table}

\begin{tabular}{|r|c c|}
\hline 
        &Bob: $C$ & Bob: $D$
        \\ \hline
        Alice: $C$ & (3,3) & (0,5)
        \\
        Alice: $D$ & (5,0) & (1,1)\\
        \hline
\end{tabular}
\label{tab:PayOff}
\caption[caption]{Pay-off matrix for the Prisoners' Dilemma. The
first entry in the parenthesis denotes the pay-off of Alice and the
second number is Bob's pay-off. The numerical values are chosen as in
\cite{DawXX}. Referring to Eq.~(\ref{eq:PayOff}) this choice
corresponds to $r=3$ (``{\it reward}\/''), $p=1$ (``{\it
punishment}\/''), $t=5$ (``{\it temptation}\/''), and $s=0$ (``{\it
sucker's pay-off}\/'').}
\end{table}

The quantum formulation proceeds by assigning the possible outcomes of
the classical strategies $D$ and $C$ two basis vectors $|D\rangle$ and
$|C\rangle$ in the Hilbert space of a two-state system, i.e., a qubit.
At each instance, the state of the game is described by a vector in
the tensor product space which is spanned by the classical game basis
$|CC\rangle$, $|CD\rangle$, $|DC\rangle$ and $|DD\rangle$, where the
first and second entry refer to Alice's and Bob's qubit, respectively. 

\begin{figure}
\centering
	\includegraphics[width=.39\textwidth]{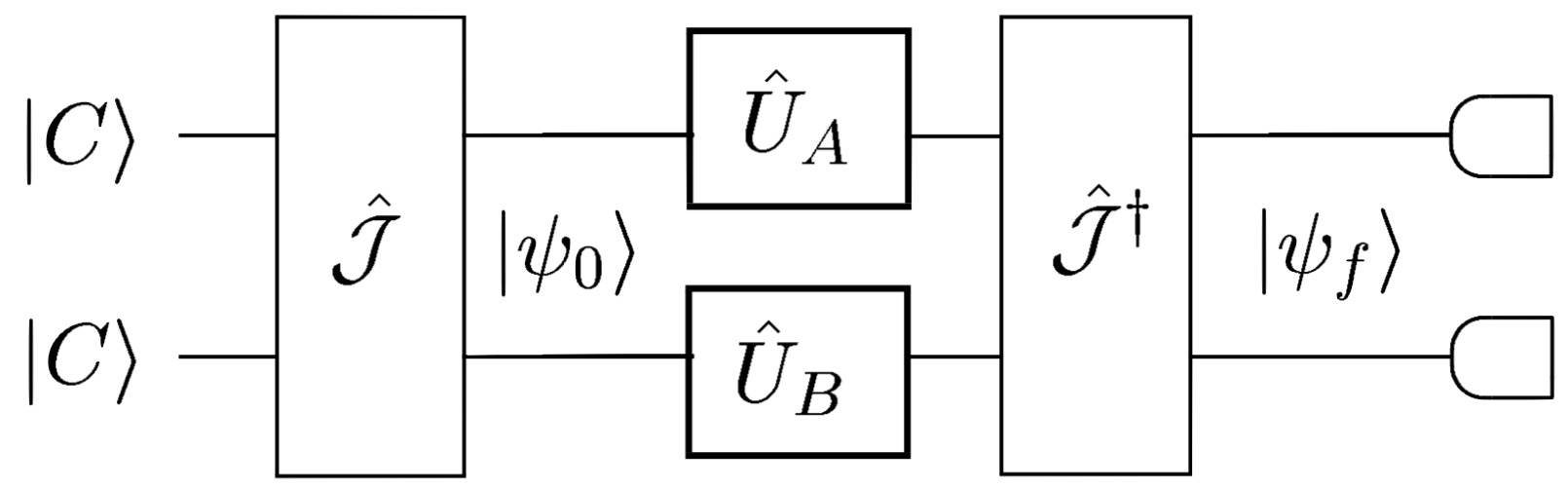}

\caption{The setup of a two player quantum game.}
\label{fig:Scheme}
\end{figure}

The board of our quantum-game is depicted in Fig.~\ref{fig:Scheme};  
it can in fact be considered a simple quantum
network \cite{Deu89} with sources, reversible one-bit and two-bit
gates and sinks. Note that the complexity is minimal in this 
implementation as the players' decisions are encoded in dichotomic 
variables.

We denote the game's
initial state vector by 
$
   |\psi_{0}\rangle = \hat{\cal J}|CC\rangle\,,
$
where $\hat{\cal J}$ is a unitary operator which is known to both
players. For fair games $\hat{\cal J}$ must be symmetric with respect
to the interchange of the two players.
The strategies are executed on the distributed pair of qubits in the
state vector  $|\psi_{0}\rangle$. Strategic moves of Alice and Bob are
associated with unitary operators $\hat{U}_{A}$ and $\hat{U}_{B}$,
respectively, which are chosen from a strategic space $S$. The
independence of the players dictates that $\hat{U}_{A}$ and
$\hat{U}_{B}$ operate exclusively on the qubits in Alice's and Bob's
possession, respectively. The strategic space $S$ may therefore be
identified with some subset of the group of unitary $2\times2$
matrices.

Having executed their moves, which leaves the game in a state vector  
$(\hat{U}_A\otimes \hat{U}_B)\hat{\cal J}|CC\rangle$, 
Alice and Bob forward their qubits for  
the final measurement which determines their pay-off.
The measurement device consists of a reversible
two-bit gate $\tilde{{\cal J}}$ which is 
followed by a pair of Stern Gerlach type detectors. The two
channels of each detector are labeled by $\sigma=C,D$.
With the proviso of subsequent
justification we set $\tilde{\cal J}=\hat{\cal J}^{\dagger}$, such
that the final state vector
$|\psi_{f}\rangle=|\psi_{f}(\hat{U}_A,\hat{U}_B)\rangle$ of the game
prior to detection is given by
\begin{equation}
   |\psi_{f} \rangle = \hat{\cal J}^{\dagger}
	\left(\hat{U}_A\otimes \hat{U}_B\right)
	\hat{\cal J}|CC\rangle\,.
\end{equation}
%
The subsequent detection yields a particular result,  
$\sigma\sigma'=CD$ say, 
and the pay-off is returned  
according to the corresponding entry of the pay-off matrix. 
Yet quantum mechanics being a fundamentally probabilistic theory, 
the only {\it strategic}\/ notion of a pay-off is the 
{\it expected}\/ pay-off.
Alice's expected pay-off is given by
\begin{equation}\label{eq:PayOff}
   {\$}_{A} = rP_{CC}+ pP_{DD} +tP_{DC}+ sP_{CD}\,,
\end{equation}
where 
$
	P_{\sigma\sigma'}=|\langle\sigma\sigma'| \psi_{f} \rangle|^{2}
$
is the joint probability that the channels $\sigma$ and $\sigma'$ of  
the Stern-Gerlach type devices will click. 
Bob's expected pay-off is obtained by interchanging $t\leftrightarrow s$ in
the last two entries (for numerical values of $r$, $p$, $t$, $s$  
see Tab.\
\ref{tab:PayOff}). Note that Alice's
expected pay-off ${\$}_A$ not only depends on her choice of strategy
$\hat{U}_A$, but also on Bob's choice $\hat{U}_B$.

It proves to be sufficient to restrict the strategic space to the
$2$-parameter set of unitary $2\times2$ matrices
\begin{equation}\label{setof}
        \hat{U}(\theta,\phi)=
        \left(
        \begin{array}{cc}
        e^{i\phi}\cos\theta/2&
        \sin\theta/2\\
        -\sin\theta/2
        &e^{-i\phi}\cos\theta/2
        \\
        \end{array}
        \right)
\end{equation}
with $0\leq \theta\leq\pi$ and $0\leq \phi\leq \pi/2$. 
To be specific, we
associate the strategy ``cooperate'' with the operator
\begin{equation}
	\hat{C}:=\hat{U}(0,0),\,\,\,\,
	\hat{C}=\left(\begin{array}{cc} 1& 0 \\ 0 & 1 \end{array}\right)\,,
\end{equation}
while the strategy ``defect'' is associated with a spin-flip,
\begin{equation}
	\hat{D}:=\hat{U}(\pi,0),\,\,\,\,
	\hat{D}=\left(\begin{array}{cc} 0 & 1 \\ -1 & 0 \end{array}\right)\,.
\end{equation}

In order to guarantee that the ordinary Prisoners' Dilemma is
faithfully represented, we impose the subsidiary conditions
\begin{equation}
   [\hat{\cal J},\hat{D}\otimes\hat{D}]=0\,,
   [\hat{\cal J},\hat{D}\otimes\hat{C}]=0\,,
   [\hat{\cal J},\hat{C}\otimes\hat{D}]=0\,.
\label{eq:correspondence}
\end{equation}
These conditions together with the identification
$\tilde{\cal J}=\hat{\cal
J}^\dagger$ imply that for any pair of strategies taken
from the subset
$S_0:= \{ \hat{U}(\theta,0)|\, \theta\in[0,\pi]\}$,
the joint probabilities $P_{\sigma\sigma'}$ 
factorize, $P_{\sigma\sigma'}=p_{A}^{(\sigma)}p_{B}^{(\sigma')}$, where  
$p^{(C)}=\cos^{2}(\theta/2)$ and $p^{(D)}=1-p^{(C)}$. 
Identifying $p^{(C)}$ with the individual preference to cooperate, 
we observe that condition (\ref{eq:correspondence}) in fact assures that
the {\it quantum}\/ Prisoners' Dilemma
entails a faithful representation of the most general 
{\it classical}\/ Prisoners' Dilemma,
where each player uses a biased coin in order to decide  
whether he or she chooses to cooperate or to defect
\cite{mixed}. Of course, the entire set of quantum strategies
is much bigger than $S_0$, and it is the quantum sector
$S \backslash S_0$ which offers additional degrees of freedom 
that can be exploited for strategic purposes.
Note that our quantization scheme applies to any two player 
binary choice symmetric game and -- due to the classical 
correspondence principle Eq.\ (\ref{eq:correspondence}) --
is to a great extent canonical.

\begin{figure}
\centering
\includegraphics[width=.4\textwidth]{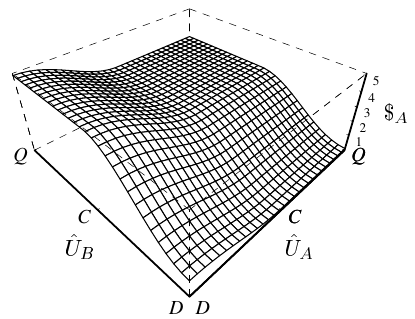}

\caption{Alice's pay-off in a separable game.
In this and the following plot we have chosen a certain
parametrization such
that the strategies $\hat{U}_A$ and $\hat{U}_B$ each depend
on a
single parameter $t\in[-1,1]$ only: we set
$\hat{U}_A=\hat{U}(t \pi, 0)$ for $t\in[0,1]$ and
$\hat{U}_A=\hat{U}(0,-t\pi/2)$ for $t\in[-1,0)$
(same for Bob).
Defection $\hat{D}$
corresponds to the value $t=1$, cooperation $\hat{C}$
to $t=0$, and $\hat{Q}$ is represented by $t=-1$.}
\label{fig:AlicesClassicalPayOff}
\end{figure}

Factoring out Abelian subgroups which yield nothing but a
reparametrization of the quantum sector of the strategic space $S$, a
solution of Eq.~(\ref{eq:correspondence}) is given by
\begin{equation}
   \hat{\cal J} = \exp(- i\gamma\hat{D}\otimes\hat{D}/2)\,,
\end{equation}
where $\gamma\in[0,\pi/2]$ is a real parameter. In fact, $\gamma$ is a
measure for the game's entanglement. 
For a separable game 
$\gamma=0$, and 
the joint probabilities $P_{\sigma\sigma'}$ factorize
for all possible pairs of strategies $\hat{U}_A, \hat{U}_B$.
Fig.~\ref{fig:AlicesClassicalPayOff}
shows Alice's expected pay-off for $\gamma=0$. 
As can be seen in this
figure, for any of Bob's choices $\hat{U}_B$ Alice's pay-off is
maximized if she chooses to play $\hat{D}$. The game being  
symmetric, the same holds for Bob and $\hat{D}\otimes\hat{D}$ is the  
equilibrium in dominant strategies. Indeed, separable games
do not display any features which go beyond the
classical game. 
	
\begin{figure}
\centering
\includegraphics[width=.4\textwidth]{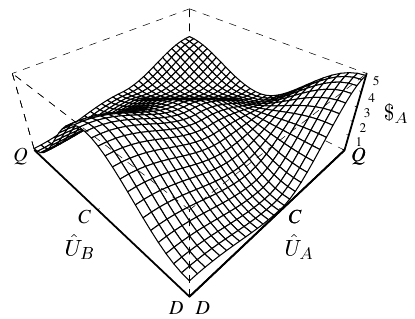}

\caption{Alice's pay-off for a maximally entangled game.
The parametrization is chosen as in  
Fig.~\ref{fig:AlicesClassicalPayOff}.} 
\label{fig:AlicesQuantumPayOff}
\end{figure}

The situation is entirely different for a maximally entangled game  
$\gamma=\pi/2$. 
Here, pairs of strategies exist which have no  
counterpart in the classical domain, yet by virtue of  
Eq.~(\ref{eq:correspondence}) the game behaves completely classical  
if both players decide to play $\phi=0$.
For example, 
$P_{CC}=|\cos(\phi_A+\phi_B)\cos(\theta_A/2)\cos(\theta_B/2)|^2$ 
factorizes on $S_0\otimes S_0$ (i.e., $\phi_A=\phi_B=0$ fixed),
but exhibits non-local correlations otherwise.
In Fig.~\ref{fig:AlicesQuantumPayOff} we depict Alice's pay-off in the
Prisoners' Dilemma as a function of the strategies $\hat{U}_{A}$,
$\hat{U}_{B}$. Assuming Bob chooses $\hat{D}$, Alice's best reply
would be
\begin{equation}
	\hat{Q}:=\hat{U}(0,\pi/2),\,\,\,\,
	\hat{Q} = \left(\begin{array}{cc} i & 0 \\ 0 &  
-i\end{array}\right)\, ,
\end{equation}
while assuming Bob plays $\hat{C}$ Alice's best strategy would be
defection $\hat{D}$.
Thus, there is no dominant strategy left for
Alice. The game being symmetric, the same holds for Bob, i.e.,
$\hat{D}\otimes \hat{D}$ is no longer an equilibrium in dominant
strategies.

Surprisingly, $\hat{D}\otimes \hat{D}$ even ceases to be a Nash
equilibrium as both players can improve by unilaterally deviating from
the strategy $\hat{D}$. However, concomitant with the disappearance
of the equilibrium $\hat{D}\otimes\hat{D}$ a new Nash equilibrium
$\hat{Q}\otimes \hat{Q}$ has emerged with pay-off
$\$_A(\hat{Q},\hat{Q})=\$_B(\hat{Q},\hat{Q})=3$. 
Indeed,
$
	\$_A(\hat{U}(\theta,\phi),\hat{Q})=
	\cos^2(\theta/2)
	\left(
		3\sin^2 \phi+ \cos^2 \phi
	\right)\leq 3
$ 
for all $\theta\in[0,\pi]$ and $\phi\in[0,\pi/2]$ and analogously
$\$_B(\hat{Q},\hat{U}_B)\leq \$_B(\hat{Q},\hat{Q})$ for all
$\hat{U}_B\in S$ such that no player can gain from unilaterally
deviating from $\hat{Q}\otimes\hat{Q}$. It can be shown 
that $\hat{Q}\otimes\hat{Q}$ is a unique equilibrium, that is,
rational reasoning dictates both players to play $\hat{Q}$ as their
optimal strategy.

It is interesting to see that $\hat{Q}\otimes \hat{Q}$ has the
property to be {\it Pareto optimal}\/ \cite{Mye91}, that is, by
deviating from this pair of strategies it is not possible to increase
the pay-off of one player without lessening the pay-off of the other
player. In the classical game only mutual cooperation is Pareto
optimal, but it is not an equilibrium solution. One could say that by
allowing for quantum strategies the players escape the dilemma
\cite{General}.

The alert reader may object that -- very much like any quantum  
mechanical system can be simulated on a classical computer -- the  
quantum game proposed here can be played by purely classical means.  
For instance, Alice and Bob each may communicate their choice of  
angles to the judge using ordinary telephone lines. The judge computes  
the values $P_{\sigma\sigma'}$, tosses a four-sided coin which is  
biased on these values, and returns the pay-off according to the  
outcome of the experiment. While such an implementation yields
the proper pay-off in this scenario four real numbers have 
to be transmitted. This contrasts 
most dramatically with our quantum mechanical model which is
more economical as far as communication resources are
concerned. Moreover, any  
local hidden variable model of the physical scheme presented
here predicts inequalities for $P_{\sigma\sigma'}$,
as functions of the four angles $\theta_A$, $\theta_B$,
$\phi_A$, and $\phi_B$, which are violated by the above expressions
for the expected pay-off. 
We conclude that in an environment with 
limited resources, it is only quantum mechanics which allows for an  
implementation of the game presented here.

\begin{figure}
\begin{center}
\hbox{

\vbox{\hsize=4.4cm (a)

\includegraphics[width=.2\textwidth]{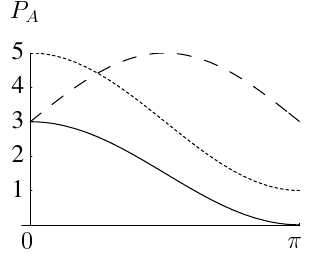}

\vspace*{-0.3cm}
$\theta$}

\vbox{\hsize=4.4cm (b)

\includegraphics[width=.2\textwidth]{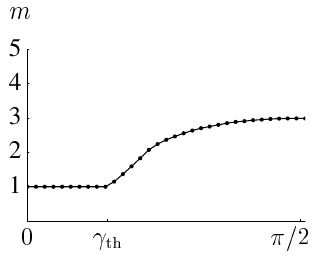}

\vspace*{-0.3cm}
$\gamma$
}

}
\end{center}
\caption{Quantum versus classical strategies: (a)
Alice's pay-off as a function of $\theta$ when
Bob plays $\hat{U}(\theta,0)$ ($\hat{U}(0,0)=\hat{C}$
and $\hat{U}(\pi,0)=\hat{D}$)
and
Alice chooses $\hat{C}$ (solid line), $\hat{D}$ (dots) 
or $\hat{M}$  
(dashes).
(b)
The expected pay-off Alice can always attain
in an unfair game as a function of the entanglement parameter $\gamma$.}
\label{fig:Unfair}
\end{figure}

So far we have considered fair games where both players had access to
a common strategic space. What happens when we introduce an unfair
situation: Alice may use a quantum strategy, i.e., her strategic space
is still $S$, while Bob is restricted to apply only ``classical
strategies'' characterized by $\phi_B=0$? 
In this case Alice is well advised to play
\begin{equation}
	\hat{M}=\hat{U}(\pi/2, \pi/2),\,\,\,\,
	\hat{M}=\frac{1}{\sqrt{2}}
	\left(\begin{array}{cc} i& -1 \\ 1 & -i \end{array}
	\right)\,,
\end{equation}
(the ``miracle move''), giving her at least reward $r=3$ as pay-off,
since $\$_A(\hat{M},\hat{U}(\theta,0))\geq3$ for any
$\theta\in[0,\pi]$, leaving Bob with
$\$_B(\hat{M},\hat{U}(\theta,0))\leq 1/2$ (see Fig.~\ref{fig:Unfair}
(a)).  Hence if in an unfair game Alice can be sure that Bob plays
$\hat{U}(\theta,0)$, she may choose ``Always-$\hat{M}$'' as her
preferred strategy in an iterated game. This certainly out-performs
{\it tit-for-tat}\/, but one must keep in mind that the assumed
asymmetry is essential for this argument.

It is moreover interesting to investigate how Alice's advantage in an
unfair game depends on the degree of entanglement of the initial state vector 
$|\psi_0\rangle$. The minimal expected pay-off $m$ Alice can always
attain by choosing an appropriate strategy $U_A$ is given by
\begin{equation}\label{eq:4}
        m=\max_{\hat{U}_A\in{S}}\min_{\hat{U}_B
        \in \{\hat{C},\hat{D}\}
        }
	\$_A(\hat{U}_A,\hat{U}_B).
\end{equation}
Alice will not settle for anything less than this quantity.
Considering $m$ a function of the entanglement parameter
$\gamma\in[0,\pi/2]$ it is clear that $m(0)=1$ (since in this case the
dominant strategy $\hat{D}$ is the optimal choice) while for maximal
entanglement we find $m(\pi/2)=3$ which is achieved by playing
$\hat{M}$.  Fig. ~\ref{fig:Unfair} (b) shows $m$ as function of the
entanglement parameter $\gamma$.  We observe that $m$ is in fact a
monotone increasing function of $\gamma$, and the maximal advantage is
only accessible for maximal entanglement.  Furthermore, Alice should
deviate from the strategy $\hat{D}$ if and only if the degree of
entanglement exceeds a certain threshold value $\gamma_{\rm
th}=\arcsin(1/\sqrt{5})\approx 0.464$. The observed threshold
behavior is in fact reminiscent of a first order phase transition in
Alice's optimal strategy: at the threshold she should discontinuously
change her strategy from $\hat{D}$ to $\hat{Q}$.

Summarizing, we have demonstrated that novel features emerge if
classical games like the Prisoners' Dilemma are extended into the
quantum domain.  We have 
introduced a correspondence principle which guarantees
that the performance of a classical game and its quantum extension
can be compared in an unbiased manner. 
Very much like in quantum cryptography and
computation, we have found superior performance of the quantum
strategies if entanglement is present.

This research was triggered by an inspiring talk of Artur Ekert on
quantum computation. We also acknowledge fruitful discussions with
S.~M.\ Barnett,
C.~H.\ Bennett,
R.\ Dum,
T.\ Felbinger,
P.~L.\ Knight,
H.-K.\ Lo,
M.~B.\ Plenio,
A.\ Sanpera,
and
P.\ Zanardi. This work was supported by the DFG.


\begin{references}
\bibitem{Footnote1}
        Alice's reasoning goes as follows: ``If Bob
        cooperates my pay-off will be maximal if and only if I  
	defect. If,
        on the other hand, Bob defects, my pay-off will again be  
	maximal if
        and only if I defect. Hence I shall defect.''.
\end{references}
\end{document}